\documentclass[a4paper,11pt]{article}

\usepackage{jheppub}

\usepackage[utf8]{inputenc}
\usepackage{amsmath}
\usepackage{graphicx}
\usepackage{color}
\usepackage[dvipsnames]{xcolor}
\usepackage[version=3]{mhchem}

\definecolor{nicegreen}{rgb}{0.1,0.5,0.1}
\definecolor{darkblue}{rgb}{0.15, 0.2, .85}
\definecolor{darkgreen}{rgb}{0.1,0,0.3}
\definecolor{darkred}{rgb}{0.6,0,0}

\newcommand{\nue}{\nu_e}
\newcommand{\nuebar}{\overline{\nu}_e}

\newcommand{\mdm}{m_\chi}
\newcommand{\Ekin}{E_{\rm kin}}

\newcommand{\Msun}{M_\odot}

\newcommand{\Mmin}{M_{\rm min}}

\newcommand{\Javg}{J_{\Delta\Omega}}

\newcommand{\MeV}{\, {\rm MeV}}

\newcommand{\cm}{\, {\rm cm}}
\newcommand{\s}{\, {\rm s}}

\newcommand{\GENIE}{{\tt GENIE }}

\newcommand{\ROOT}{{\tt ROOT }}

\newcommand{\Omegam}{\Omega_{m,0}}
\newcommand{\OmegaL}{\Omega_{\Lambda,0}}
\newcommand{\OmegaDM}{\Omega_{\rm DM,0}}
\newcommand{\sigmav}{\langle \sigma v \rangle}

\newcommand{\zmax}{z_{\rm max}}

\definecolor{ShamrockGreen}{rgb}{0.0, 0.62, 0.38}

\newcommand{\mjdcomm}[1]{{\color{ShamrockGreen}  #1}}

\title{Dark Matter Pollution in the Diffuse Supernova Neutrino Background}

\author[a]{Nicole F. Bell,}
\author[a]{Matthew J. Dolan,}
\author[b]{Sandra Robles}
\affiliation[a]{ARC Centre of Excellence for Dark Matter Particle Physics, \\
School of Physics, The University of Melbourne, Victoria 3010, Australia}
\affiliation[b]{Theoretical Particle Physics and Cosmology Group, Department of Physics, King's College London, Strand, London, WC2R 2LS, UK}
\emailAdd{n.bell@unimelb.edu.au}
\emailAdd{matthew.dolan@unimelb.edu.au}
\emailAdd{sandra.robles@kcl.ac.uk}

\abstract{
The Hyper-Kamiokande (HyperK) experiment is expected to precisely measure the Diffuse Supernova Neutrino Background (DSNB). This requires that the backgrounds in the relevant energy range are well understood. One possible background that has not been considered thus far is the annihilation of low-mass dark matter (DM) to neutrinos. We conduct simulations of the DSNB signal and backgrounds in HyperK, and quantify the extent to which DM annihilation products can pollute the DSNB signal. We find that the presence of DM could affect the determination of the correct values of parameters of interest for DSNB physics, such as effective neutrino temperatures and star formation rates. While this opens the possibility of simultaneously characterising the DNSB and discovering dark matter via indirect detection, we argue that it would be hard to disentangle the two contributions due to the lack of angular information available at low energies.
}

\begin{document}
\hfill KCL-PH-TH/2022-32
\maketitle

%%%%%%%%%%%%%%%%%%%%%%%%%%%%%%%%%%%%%%%%%%
\section{Introduction}
\label{sec:intro}
%%%%%%%%%%%%%%%%%%%%%%%%%%%%%%%%%%%%%%%%%%%%%%%%%%%%%%%%%%%%%%

The next decade will see the commissioning of a new generation of neutrino telescopes. These include DUNE, JUNO and Hyper-Kamiokande (HyperK). Collectively, these experiments will address a number of outstanding issues in neutrino physics and astrophysics. One keenly anticipated measurement is that of the Diffuse Supernova Neutrino Background (DSNB)~\cite{Beacom:2010kk,Mirizzi:2015eza}, also known as Supernova Relic Neutrinos (SRNs). The DSNB is the isotropic flux of neutrinos emitted by all past core-collapse supernovae in the Universe. These neutrinos have $\cal{O}(\textrm{10 MeV})$ energies, with a quasi-thermal spectrum determined by the details of the neutrino emission from the protoneutron star, and a flux which is set by the cosmic supernova rate and hence the star formation rate.

Precise characterisation of supernova energy spectra and flavour composition would reveal a wealth of information about supernova progenitors, the core-collapse process, neutrino mixing effects in the dense proto-neutron star medium, and possible exotic neutrino physics. While a Galactic core-collapse supernova would provide an abundance of events in neutrino telescopes, allowing a precise determination of these parameters, the Galactic supernova rate is low, with a rate of order a few per century. The DSNB, on the other hand, is a steady-state flux of extragalactic supernova neutrinos, with an expected event rate in next generation neutrino telescopes large enough to begin to probe the questions above. In addition, measurement of the DSNB will deliver information on cosmic quantities such as the cosmic core-collapse rate and the fraction of black-hole forming supernova~\cite{Moller:2018kpn,DeGouvea:2020ang}. A recent analysis~\cite{Li:2022myd} estimates a $3\sigma$ detection of the DSNB will be achieved in  2026, reaching $5\sigma$ in 2031 through a combination of Super-Kamiokande (SuperK) and JUNO data. The JUNO Collaboration, which is due to commence in 2023, estimates that JUNO will reach $3\sigma$ significance after three years runtime, and over $5\sigma$ after 10 years~\cite{JUNO:2022lpc}. Precise measurements from the experiments DUNE, JUNO and HyperK will arrive in the 2030s.

The expected event rates for the DSNB are low. Therefore a precise understanding of the backgrounds will be critical in achieving a robust detection and in accurately characterising the spectrum. The main detection channel is the inverse beta decay of electron-antineutrinos, $\overline{\nu}_e  + p \rightarrow e^+ + n$.
The DSNB flux peaks around 5 MeV depending on the precise model
and parameters that are assumed, and yields a non-zero expected number of events up to energies of around 50~MeV (for a 10-year exposure at HyperK).  However, there are insurmountable backgrounds at the lowest energies where the DSNB event rate would peak. The reactor $\overline{\nu}_e$ background is overwhelming below about 10 MeV, and cosmic-ray induced spallation events also swamp the DSNB signals at low energies.\footnote{Solar neutrinos contribute a large background at energies below about 20 MeV, but can be reduced to negligible levels via neutron-tagging~\cite{Super-Kamiokande:2021jaq} or directional information.} At higher energies, atmospheric neutrinos contribute a significant background. Consequently, the most recent SuperK analysis takes place within a window of 9 to 31~MeV~\cite{Super-Kamiokande:2021jaq}. 
Because HyperK will be at a shallower site than SuperK, providing less shielding from cosmic rays, spallation backgrounds will be larger. The HyperK design report~\cite{Abe:2018uyc} thus assumes a smaller analysis window of 16~MeV to 30~MeV~\footnote{However, these backgrounds and removal methods are becoming better understood both theoretically and experimentally~\cite{Li:2014sea,Li:2015kpa,Li:2015lxa,Super-Kamiokande:2021snn} so it may be possible to lower the edge of the analysis window. That would also assist in discriminating between dark matter and the DSNB.}.

In this paper we pose the question: can neutrinos from dark matter (DM) annihilation contribute a background for DSNB searches? If so, what impact would this have on extracting the DSNB signal? In general, the existence of dark matter as a background in the DSNB energy range could cause incorrect inferences to be made about key astrophysical quantities. Is there a way to reliably disentangle DSNB and DM signals? 

The possibility of DM pollution of a DSNB signal was not relevant for previous DSNB analyses.  This is because existing searches have not had the sensitivity to detect either the DSNB, or the neutrinos produced via the annihilation of low mass dark matter with a thermal relic cross section. This includes the recent SuperK analysis that currently sets the most sensitive limit on the DSNB $\overline{\nu}_e$ flux~\cite{Super-Kamiokande:2021jaq}. However, we have shown in previous work~\cite{Bell:2020rkw} that HyperK will have sensitivity to thermal relic dark matter annihilating into neutrinos in precisely this mass range. In that work we assumed that the DSNB was a known (and fixed) component of the background in the DM search. In this work we assume that light DM is an unknown background within the DSNB analysis window.
 
We show that current DSNB analysis strategies are susceptible to being polluted by low mass dark matter annihilation. However, as we will discuss, it will be difficult to mitigate against the presence of DM within the DSNB analysis window, due to the lack of angular information in low-energy inverse beta decay events. The paper is structured as follows: In Section~\ref{sec:dm} we detail our calculations of the DSNB and DM signals and atmospheric neutrino backgrounds. In Section~\ref{sec:results} we illustrate the impact that DM can have on DSNB analyses and model discrimination, and demonstrate how angular information can be used to mitigate this effect. Our discussion and outlook can be found in Section~\ref{sec:summary}.

%%%%%%%%%%%%%%%%%%%%%%%%%%%%%%%%%%%%%%%%%%%%%%%%%5
\section{Signal and Background Calculation}
\label{sec:dm}
%%%%%%%%%%%%%%%%%%%%%%%%%%%%%%%%%%%%%%%%%%%%%%%%5

In this section we summarise our calculations of the signal and background rates. The Galactic dark matter and background calculations generally follow our previous work in~\cite{Bell:2020rkw}, to which the reader is referred for details. We focus primarily on what is new to this work, namely the extragalactic dark matter component and the DSNB. We note that the HyperK Collaboration has slightly changed the planned detector dimensions since the publication of our previous work~\cite{Bell:2020rkw,Bell:2021esh}. The original Design Report~\cite{Abe:2018uyc} specified a diameter of $74$~m and height of $60$~m. These parameters are $68$~m and $71$~m, respectively, in the new design\footnote{\url{http://www.hyper-k.org/en/detector/detector-detail.html}}. This changes the cross-sectional profile depending on the incidence angle of the incoming neutrinos, although the total volume is unchanged and the fiducial volume changes only marginally from 187~kton to 188~kton. We use the updated parameters in this work.

Events are generated with \GENIE v3.0.4a \cite{Andreopoulos:2009rq} plus atmospheric neutrino flux and detector geometry drivers~\cite{Andreopoulos:2015wxa}, for the tune G18\_10,  using the detector geometry implemented with the \ROOT geometry package \cite{Brun:1997pa}. For parton distribution functions we use the GRV98 LO PDFs~\cite{Gluck:1998xa} provided via LHAPDF~\cite{Whalley:2005nh}.

%%%%%%%%%%%%%%%%%%%%%%%%%%%%%%%%%%%%%%%%5
\subsection{Background Calculation}
\label{subsec:backgroundcalc}
%%%%%%%%%%%%%%%%%%%%%%%%%%%%%%%%%%%%%%%%%%%%%%%5

For energies above 100~MeV we use the atmospheric 4D neutrino flux at the Kamioka site with mountain over the detector from HKKM11~\cite{Honda:2011nf}. Below 100~MeV we use the expected flux at Kamioka, integrated over solid angle, calculated using FLUKA~\cite{Battistoni:2005pd}. We assume that the angular distribution for all the energy bins of the FLUKA flux is the same as that of the HKKM11 flux at 100~MeV. For the HKKM11 flux we generate events for 5 years of solar minimum and 5 years of solar maximum, and for FLUKA for 10 years of solar average. Although our primary region of interest in this paper is at neutrino energies below 100~MeV, neutrinos with energies $E_{\nu}\geq 100$~MeV do make a contribution to the background once we bin in the final-state lepton kinetic energy, $\Ekin$~\cite{Bell:2020rkw}.

Neutrino oscillations are computed with \texttt{nuCraft}~\cite{Wallraff:2014qka} using the Preliminary Earth Reference Model~\cite{Dziewonski:1981xy} and neutrino parameters from the Particle Data Group~\cite{Zyla:2020zbs}, assuming a normal mass ordering as in~\cite{Bell:2021esh}.

DSNB searches at water Cherenkov detectors rely on the inverse beta-decay process, $\bar{\nu}_e + p \to e^+ + n$.  This process contains a final state neutron, while many of the background processes do not. Therefore, neutron-tagging provides a powerful method of background rejection. The method of neutron tagging currently used at SuperK relies on direct neutron rapid thermalisation, after which they can be captured by hydrogen atoms (the cross-section for capture on oxygen is much smaller). This occurs via the reaction $n+p\to d + \gamma(2.2~\rm{MeV})$. This 2.2~MeV photon is then used to tag the neutron via a delayed coincidence trigger~\cite{Super-Kamiokande:2008mmn}. Neutron tagging has been used in DSNB searches since the SuperK-IV analysis~\cite{Zhang:2013tua}. A second method relies on the addition of gadolinium (Gd) salt to the water in the detector~\cite{Beacom:2003nk}. The Gd capture cross-section is orders of magnitude larger than for hydrogen, and the excited Gd nucleus then emits 8~MeV of photons in a cascade which is easily detectable.

In discussing the backgrounds relevant for DSNB detection, it is important to make a distinction between the neutrino energy, $E_\nu$, and the quantity that is actually measured. The latter is the visible energy in the detector, $E_{vis}$, which is related to the energy of the relativistic leptons produced in the neutrino interactions, $E_{kin}$. Importantly, note that neutrinos with energies as high as about 300 MeV can interact to produce events with $E_{vis}$ low enough to fall in the DSNB analysis window.  In particular, atmospheric $\nu_\mu$ and $\overline{\nu}_\mu$ can interact to produce non relativistic muons and antimuons that are below the threshold to radiate Cherenkov light. The decay of these ``invisible muons" to $e^+$ or $e^-$ contributes a key background in the DSNB energy range.\footnote{We take the ``invisible muon" events from the HyperK Design Report~\cite{Abe:2018uyc}.}  However, the absence of a final state neutron allows neutron-tagging to significantly reduce this background. Likewise, charged current interactions of atmospheric $\nu_e$, namely $\nu_e + n \to e^- +p$, can also be rejected due to the absence of a final state neutron. (While this process can yield secondary neutrons, they can be neglected in the DSNB energy range.) We therefore ignore charged current atmospheric $\nue$ events in this study. We also assume that we can ignore backgrounds from neutral current interactions~\cite{Moller:2018kpn,Maksimovic:2021dmz}.  Finally, neutron tagging also leads to a significant reduction of spallation backgrounds.

The remaining backgrounds dictate the upper and lower thresholds for the DSNB analysis.  At low energy, those remaining backgrounds are reactor antineutrinos and spallation events, with the latter setting the lower threshold for DSNB searches. At HyperK, this is expected to be around 16~MeV~\cite{Abe:2018uyc}, which is what we adopt in our analysis.  At higher energy, the key remaining background is the low energy tail of the atmospheric $\overline{\nu}_e$ flux.

%%%%%%%%%%%%%%%%%%%%%%%%%%%%%%%%%%%%%%%%%%%%%%%%%%%5
\subsection{Dark Matter Calculation}
%%%%%%%%%%%%%%%%%%%%%%%%%%%%%%%%%%%%%%%%%%%%%%%%

The primary source of neutrinos from dark matter annihilation is the Galactic centre component. 
We refer to our previous work~\cite{Bell:2020rkw} for how we calculate this flux. Note that the astrophysical J-factor, which is defined in Galactic coordinates, provides the angular dependence of the signal at the detector location, after properly performing the transformation from Galactic  to horizontal $(a,z)$ coordinates. This calculation involves tracking the position of the Galactic centre in the sky. To account for this uncertainty we have averaged the J-factor over a 24 hours period as detailed in Appendix B of ref.~\cite{Bell:2020rkw}.
The differential flux of neutrinos produced from Galactic dark matter annihilation is then 
\begin{equation}
\dfrac{{d\Phi_\nu}_{\Delta\Omega}}{dE_\nu} = \frac{\sigmav}{8\pi \mdm^2}  \Javg(a,z) \dfrac{dN_\nu}{dE_\nu} \, ,
\label{eq:dmflux}
\end{equation}
where $\langle \sigma v \rangle$ is the thermally averaged annihilation cross section, $m_\chi$ is the DM mass, $J_{\Delta\Omega}(a,z)$ is the angle-averaged J-factor, and $\frac{dN_{\nu}}{dE_{\nu}}$ is the neutrino differential energy spectrum.

A secondary neutrino component arises from extragalactic dark matter annihilation. The diffuse neutrino flux from extragalactic DM annihilation is given by~\cite{Beacom:2006tt}
\begin{equation}
\dfrac{d\Phi_\nu}{dE_\nu} = \frac{\sigmav}{2}\frac{c}{4\pi H_0}\frac{\OmegaDM^2\rho_{c,0}^2}{\mdm^2}\int_0^{z_{up}} dz \frac{\Delta^2(z)}{h(z)} \dfrac{dN_\nu(E'_\nu)}{dE'_\nu},
\end{equation}
where $H_0$ is the Hubble constant, $h(z)=\sqrt{\Omegam (1+z)^3+\OmegaL}$ and $\rho_{c,0}$ is the critical density of the Universe at $z=0$. The quantities $\OmegaDM$, $\Omegam$ and $\OmegaL$ are, respectively, the dark matter, matter (DM and baryons) and dark energy densities (in units of $\rho_{c,0}$),
\begin{equation}
\dfrac{dN_\nu(E'_\nu)}{dE'_\nu} = \frac{2}{3E_\nu}\delta\left[{z-\left(\frac{\mdm}{E_\nu}-1\right)}\right], 
\end{equation}
is the neutrino spectrum ($\nu+\bar{\nu}$). The factor $\Delta^2(z)$, which accounts for the enhancement to the annihilation rate due to the DM clustering in halos, is defined in Table~\ref{tab:DeltaFactor}, where the values of $f_0$ for the NFW, Moore, and Kravtsov profiles are $5\times 10^4$, $5\times 10^5$, and $2\times 10^4$, respectively.
The halo boost factor $G(z)$
in this table, given in ref.~\cite{Lopez-Honorez:2013cua}, is calculated by summing the contribution of all halos, i.e., a single halo contribution is weighted by the halo mass function (HMF), $dn/dM$:
\begin{equation}
G(z) = \frac{1}{\OmegaDM^2 \rho_{c,0}^2}\frac{1}{(1+z)^6}\int_{\Mmin}^\infty dM \dfrac{dn(M,z)}{dM} \int_0^{R_{200}} 4\pi r^2 \rho_{\rm DM}^2(r).
\end{equation}
Halos are assumed to be spherical and follow an NFW profile, with radius $R_{200}$, i.e. the radius 
at which the mean matter density enclosed within is 200 times the critical density of the Universe at redshift $z$ $\rho_c(z)$. The second integral is computed as a function of the concentration parameter $c_{200}(M,z)$ using
the fits from the MultiDark/BigBolshoi simulations~\cite{Prada:2012}. The HMF is calculated using the parametrizations in refs.~\cite{Watson:2013,Tinker:2008ff,Press:1974,Bond:1990iw}. The integral over the halo mass is evaluated for two extreme cases of the minimum halo mass: $\Mmin=10^{-3}\Msun$ and $\Mmin=10^{-9}\Msun$ as in ref.~\cite{Arguelles:2019ouk}, since this value is not well constrained. 

\begin{figure}[t]
    \centering
    \includegraphics[width=\textwidth]{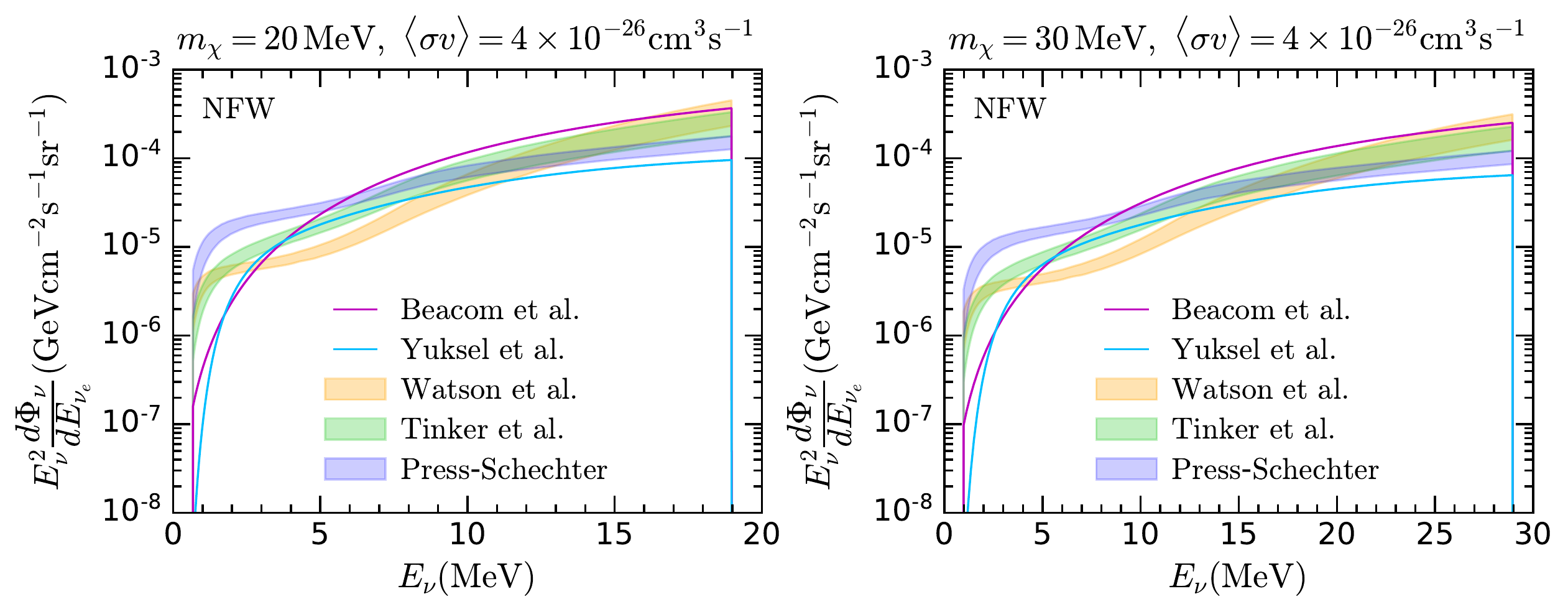}   
    \caption{Neutrino flux from extragalactic DM annihilation for $\sigmav=4\times10^{-26}\cm^3\s^{-1}$, assuming an NFW profile, $\mdm=20\MeV$ (left) and $\mdm=30\MeV$ (right), and different parametrizations of the factor $\Delta^2(z)$ in Table~\ref{tab:DeltaFactor}. The regions shaded in orange, green and blue correspond to the computation in ref.~\cite{Arguelles:2019ouk}, using the HMF in refs.~\cite{Watson:2013}, \cite{Tinker:2008ff} and \cite{Press:1974,Bond:1990iw},  respectively to calculate the boost factor $G(z)$~\cite{Lopez-Honorez:2013cua}. }
    \label{fig:extragalflux}
\end{figure}

In Fig.~\ref{fig:extragalflux} we show the neutrino flux from extragalactic dark matter annihilation, assuming an NFW halo profile and dark matter masses of $\mdm=20$~MeV (left panel) and $\mdm=30$~MeV (right panel). The lines correspond to different parametrisations of the enhancement factor $\Delta^2(z)$ defined in Table~\ref{tab:DeltaFactor}.
The solid magenta line is the calculation of~\cite{Beacom:2006tt}, and the solid blue line that of~\cite{Yuksel:2007ac}. The orange, green, and blue shaded regions correspond to the calculation of the boost factor $G(z)$~\cite{Lopez-Honorez:2013cua} in ref.~\cite{Arguelles:2019ouk} using the halo mass function of refs.~\cite{Watson:2013}, \cite{Tinker:2008ff} and~\cite{Press:1974,Bond:1990iw} respectively.

\begin{figure}[t] 
\centering
\includegraphics[width=\textwidth]{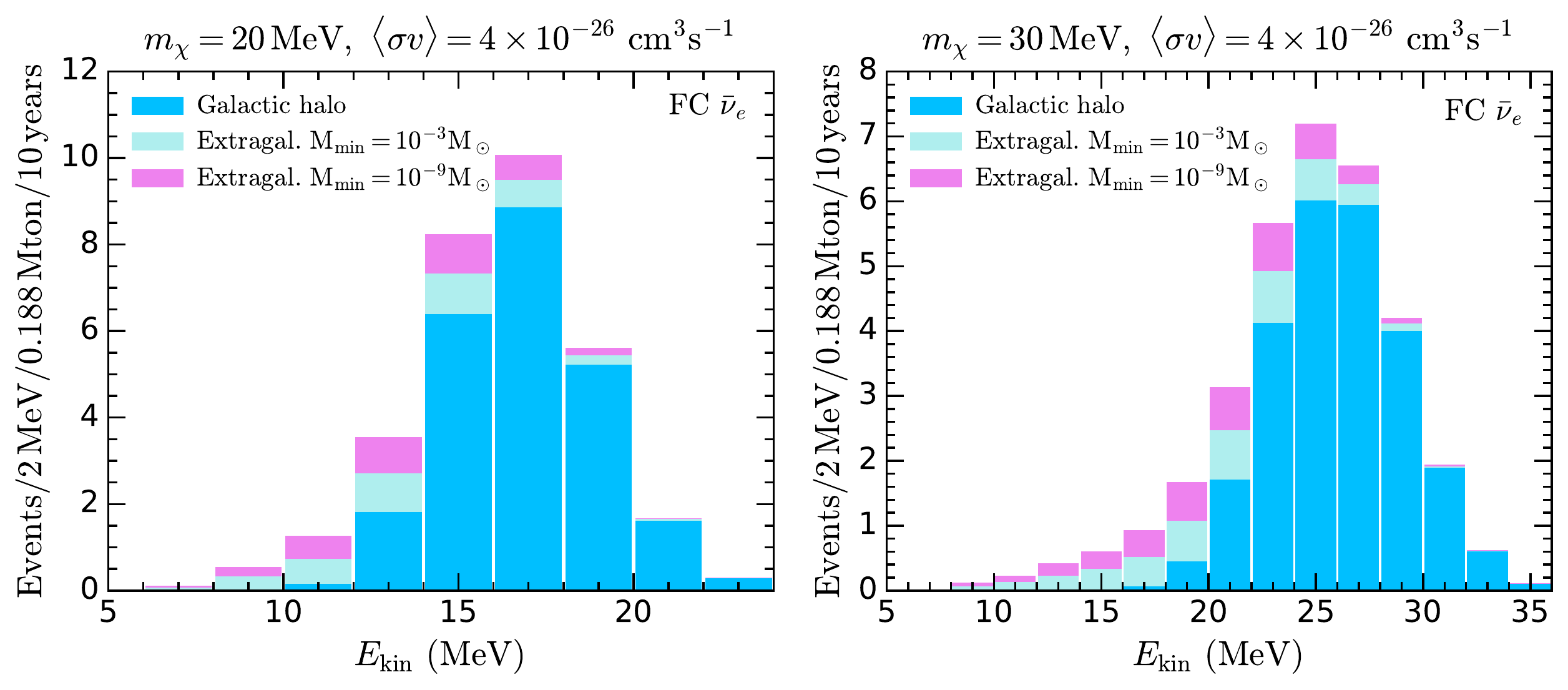} 
\caption{Expected DM induced $\nuebar$ events at HyperK as a function of the kinetic energy of the positron, calculated using the \GENIE tune G18\_10a, for 10 years livetime,  $\mdm=20\MeV$  (left) and  $\mdm=30\MeV$ (right) and $\sigmav=4\times10^{-26}\cm^3\s^{-1}$. Events from DM annihilation in the GC are shown in blue and those from the extragalactic contribution calculated using the boost factor $G(z)$, with HMF from Watson et al.~\cite{Watson:2013} and $\Mmin=10^{-3}\Msun$ are shown  in cyan. Additional events obtained with the same HMF  and $\Mmin=10^{-9}\Msun$,  which gives the most sizeable contribution, are depicted in magenta.  
} 
\label{fig:HK_DM2}
\end{figure}

The total spectrum of dark matter induced $\nuebar$ events at HyperK for 10 years running time is shown in Fig.~\ref{fig:HK_DM2}, for $\mdm=20$~MeV (left panel) and 30~MeV (right panel). The events are shown as a function of the positron kinetic energy $\Ekin$. The annihilation cross-section is taken to be $\langle \sigma v \rangle=4\times 10^{-26}\rm{cm}^3\rm{s}^{-1}$ which yields the correct relic density for those DM masses.
Events from DM annihilation in the Galactic centre are shown in blue, and exhibit a peak below, yet near, the DM mass. These events correspond to scattering off hydrogen. A smaller peak due to scattering off oxygen is below the detector threshold, for both choices of DM mass. The extragalactic contribution uses the boost factor $G(z)$ and the halo mass function from ref.~\cite{Watson:2013} with minimum halo mass $M_{\rm{min}}=10^{-3 }\Msun$ (cyan), with additional events in magenta for $M_{\rm{min}}=10^{-9}\Msun$. The extragalactic events lie primarily in vicinity of the lower energy tail of the GC signal, due to redshifting of the neutrino energy.

\begin{table}[t]
    \centering
    \begin{tabular}{|c|c|}
    \hline
    $\Delta^2(z)$  & Ref. \\
    \hline
    $2\times10^5$ & \cite{Beacom:2006tt}\\
     $f(z) = f_0 (1+z)^3 10^{0.9[\exp{(-0.9z)}-1]-0.16z}$ & \cite{Yuksel:2007ac} \\
     $(1 + G(z)) (1+z)^3$ & \cite{Arguelles:2019ouk}\\
     \hline
    \end{tabular}
    \caption{Factor that accounts for the enhancement to the annihilation rate due to the DM clustering in halos which is a function of redshift $z$, assuming an NFW halo density profile. The $f(z)$ fitting is also given for Moore and Kravtsov profiles with different $f_0$ (values in the main text). }
    \label{tab:DeltaFactor}
\end{table}

%%%%%%%%%%%%%%%%%%%%%%%%%%%%%%%%%%%%%%%%%%%%%%%%%%%%
\subsection{DSNB Calculation}
%%%%%%%%%%%%%%%%%%%%%%%%%%%%%%%%%%%%%%%%%%%%%%%%

The Diffuse Supernova Neutrino Background (DSNB) flux is calculated by integrating over time the rate of core-collapse supernovae (SNe) $R_{CCSN}(z)$, multiplied by the neutrino emission per SN, $dN/dE$, and including the appropriate redshift factor, 
\begin{equation}
\dfrac{d\Phi_{\nuebar}}{dE_{\nuebar}}=\frac{c}{H_0}\int_0^{\zmax} \frac{R_{CCSN}(z)}{\sqrt{\Omegam(1+z)^3+\OmegaL}} \dfrac{dN_{\nuebar}}{dE_{\nuebar}^{'}}(E_{\nuebar}^{'}) dz,
\label{eq:DSNBflux}
\end{equation}
where $E_{\nuebar}^{'}=E_{\nuebar}(1+z)$ and $\zmax=5$. To calculate the SN rate, two astrophysical inputs are required:  an initial mass function (IMF), $\xi$, and the star formation rate (SFR), $ \dot{\rho}_\star$. Both are constrained by observations. 

The IMF is a key ingredient for galaxy formation studies. It specifies the mass distribution of a stellar population. After a burst of star formation, let there be 
\begin{equation}
dN = N_0 \xi(M) dM 
\end{equation} 
  stars with  masses in the range $M$ to $M+dM$, where $N_0$ is a normalisation constant. The parameter $\xi$ is normalised so that the total mass of new born stars is
\begin{equation}
\int M \xi(M) dM = \Msun.
 \end{equation}  
 Then $N_0$ is the number of  solar masses contained in the star formation burst. 
  Traditional choices of IMF are Salpeter \cite{Salpeter:1955it}, a single power law,  Kroupa \cite{Kroupa:2000iv}, Baldry-Glazebrook (BG)~\cite{Baldry:2003xi},  
  \begin{align}
\xi(M) =  \begin{cases}
\, M^{-\alpha_2} \quad &M>0.5 \Msun,\\
\,M^{-\alpha_1}\quad & 0.1 \Msun < M < 0.5 \Msun, 
\end{cases}
\end{align}  
 where for the Salpeter IMF $\alpha_1=\alpha_2=2.35$.  We only consider the Salpeter IMF in this work.

For the SFR density, a continuous  broken power-law is assumed \cite{Yuksel:2008cu}
 \begin{equation}
  \dot{\rho}_\star(z) =  \dot{\rho}_0\left[ (1+z)^{\alpha\eta}  + \left( \frac{1+z}{B}\right)^{\beta\eta} + \left( \frac{1+z}{C}\right)^{\gamma\eta} \right]^{1/\eta}, 
  \label{eq:SFRrho}
\end{equation}  
 where $\dot{\rho}_0$ is a normalisation constant in units $\Msun \, \rm yr^{-1} \, Mpc^{-3}$,  $\eta\simeq-10$~\cite{Yuksel:2008cu} and 
\begin{eqnarray}
B &=& (1+z_1)^{1-\alpha/\beta}, \\
C &=&  (1+z_1)^{(\beta-\alpha)/\gamma} (1+z_2)^{1-\beta/\gamma} \,.
\end{eqnarray}  
The logarithmic slopes of the low, intermediate, and high redshift regimes,  $\alpha$,  $\beta$ and $\gamma$ respectively and the redshift breaks $z_1=1$ and $z_2=4$ are obtained by fitting Hubble and gamma-ray burst data \cite{Horiuchi:2008jz}; see Table~\ref{tab:SFRfits}.

\begin{table}[t]
\centering
\begin{tabular}{|l|c|c|c|c|}
\hline
Analytic fits&  $ \dot{\rho}_0$ & $\alpha$  & $\beta$ & $\gamma$  \\
\hline
Upper & 0.0213 & 3.6 & -0.1 & -2.5 \\
Fiducial & 0.0178 & 3.4 & -0.3 & -3.5\\
Lower & 0.0142 & 3.2 & -0.5 & -4.5 \\
\hline
\end{tabular}
\caption{SFR density fits for Salpeter IMF from ref.~\cite{Horiuchi:2008jz}.  For the Kroupa and BG IMFs, $ \dot{\rho}_0$  decreases by a factor $\simeq0.66$ and $\simeq0.55$, respectively.}
\label{tab:SFRfits}
\end{table}  

The core-collapse SN rate is then 
\begin{equation}
R_{CCSN}(z) = \dot{\rho}_\star(z) \frac{\int_8^{50} \xi(M) dM}{\int_{0.1}^{100} M \xi(M) dM}.
\label{eq:SNR}
\end{equation}
It is worth noting that there is uncertainty in the lower and upper mass threshold in the numerator of Eq.~\ref{eq:SNR}, a  discussion on this point can be found in ref.~\cite{Horiuchi:2008jz}. We have taken the minimum mass for the progenitor to undergo core-collapse to be $8\Msun$. This is the white dwarf - type II SN threshold, but this limit is not well defined and depends on the composition of the star, (i.e., its metallicity). The maximum mass corresponds to the threshold for direct collapse into a black hole, which is also not well known.

\begin{figure}[t] 
\centering
\includegraphics[width=0.5\textwidth]{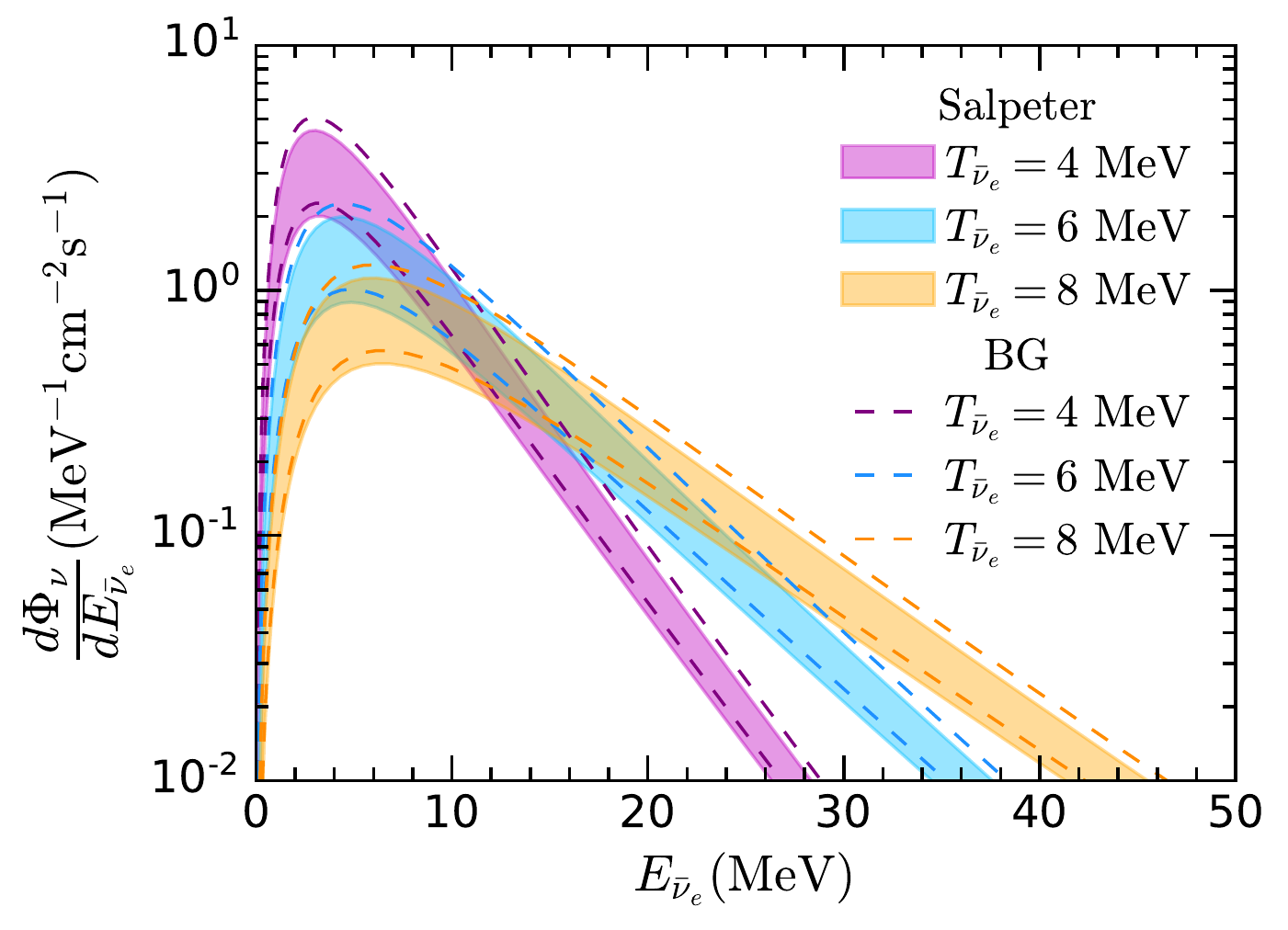}
\caption{DSNB flux for the upper and lower fits of the SFR density for the  Salpeter IMF (shaded regions) and the Baldry-Glazebrook IMF (dashed), calculated using Eq.~\ref{eq:DSNBflux}.}. 
\label{fig:diffflux}
\end{figure}

Finally, the time-integrated $\nuebar$ spectrum per supernova is well approximated by the Fermi-Dirac distribution with zero chemical potential \cite{Raffelt:1996wa,Kotake:2005zn}
\begin{equation}
\dfrac{dN_{\nuebar}}{dE_{\nuebar}^{'}}(E_{\nuebar}^{'})=\frac{120}{7\pi^4}\frac{E_\nu^{tot}}{6}\frac{E_{\nuebar}^{'2}}{T_{\nuebar}^4}\frac{1}{e^{E_{\nuebar}^{'}/T_{\nuebar}}+1}, 
\label{eq:nuebarspec}
\end{equation}
where $T_{\nuebar}$ is the effective $\nuebar$ temperature outside the star after neutrino mixing, and $E_\nu^{tot}\simeq 3\times10^{53}\,$ erg~\cite{Horiuchi:2008jz} is the total neutrino energy, considering all flavours of neutrinos and antineutrinos. 
Using Eqs.~\ref{eq:DSNBflux}, \ref{eq:SNR}, \ref{eq:SFRrho} and \ref{eq:nuebarspec}, we calculate the DSNB differential flux for the upper and lower fits in Table~\ref{tab:SFRfits} for Salpeter and BG IMFs; see Fig.~\ref{fig:diffflux}. We show the results for $T_{\nu}=4,6$ and 8~MeV in purple, blue and orange, respectively. The shaded regions correspond to the Salpeter IMF, and the dashed lines to the Baldry-Glazebrook IMF, which can be compared with Fig.~4 of ref.~\cite{Horiuchi:2008jz}. We see in Fig.~\ref{fig:diffflux} that the choice of IMF (i.e. Salpeter vs Baldry-Galzebroook) does not play a large role in the final DSNB flux. Indeed, the effect of the IMF is largely cancelled out by the SFR in the ratio in expression Eq.~\ref{eq:SNR} for the core-collapse SN rate, as has previously been noted in ref.~\cite{Horiuchi:2008jz}. We also note that there are a number of more sophisticated calculations of the DSNB flux in the literature. These take into account neutrinos from failed SN which collapse directly into black holes~\cite{Lunardini:2009ya,Nakazato:2015rya,Horiuchi:2017qja}, the impact of binary interactions such as mass transfer and mergers~\cite{Horiuchi:2020jnc}, and astrophysical uncertainties in the DSNB flux calculation~\cite{Kresse:2020nto}. While these would change the precise curves in Fig.~\ref{fig:diffflux}, we have no reason to believe they would alter our overall conclusions.

%%%%%%%%%%%%%%%%%%%%%%%%%%%%%%%%%%%%%%%%%%
\section{Results}
\label{sec:results}
%%%%%%%%%%%%%%%%%%%%%%%%%%%%%%%%%%%%%%%%%%%%%%%%%%%%%%%%%%%%%%

As discussed in section~\ref{subsec:backgroundcalc}, one important background for DSNB analyses arises from low energy muons produced by atmospheric $\nu_\mu$. When these low energy muons decay, the electron which results cannot be associated with the decay of a visible muon. These are thus known as invisible muons. Neutron tagging provides a method to suppress the invisible muon backgrounds, through the identification of neutrons in true inverse beta decay events, $\nuebar + p \to n + e^+$. We have not modelled neutron tagging in our simulations.  We instead use the distributions for the invisible muon spectrum with and without neutron tagging from the HyperK Design Report, as part of our backgrounds in the discussion in this section. The backgrounds from atmospheric $\nuebar$ are the same in both cases. We will first study the impact of dark matter in the absence of neutron tagging, showing it can affect the detection significance of the DSNB. We then study DSNB model discrimination assuming neutron tagging using Gadolinium, finding that DM can have an impact in some circumstances. 

The HyperK Design Report (DR) considers an analysis window between 16 and 30~MeV. The range below 16 MeV is dominated by spallation backgrounds, and above 30~MeV by atmospheric neutrinos.  According to the DR, after 10 years of running the expected number of DSNB events will be 70 with a statistical error of 17 events, using the DSNB flux prediction of~\cite{Ando:2002ky}. This would correspond to a $4.2\sigma$ detection of the DSNB, where the backgrounds assume the absence of neutron tagging with Gadolinium. The closest model that we consider to this is the fiducial Salpeter IMF with $T_{\nu}=6$~MeV, which  yields 75 events over 10 years, which we find corresponds to a $3.3\sigma$ significance. The definitions of the event categories and uncertainties used are the same as in our previous work~\cite{Bell:2020rkw}.

What are the implications if there is a population of neutrinos in this energy window due to dark matter annihilation?  As an example, let us consider dark matter annihilating directly into neutrinos, with the extra-galactic component determined by the Press-Schechter halo mass function with $\Mmin=10^{-3}\Msun$. We continue to assume the absence neutron tagging for our backgrounds. For DM masses larger than 30~MeV, we find no impact on the significance of the DSNB detection, since the DM signal falls mostly above the upper threshold of the analysis window. On the other hand, for DM masses in the lower range of the analysis window, the DM-induced flux adds to the DSNB flux, increasing the detection significance over the background. For example, for $\mdm=20$~MeV we find that the significance of the detection above the background rises to $7.5\sigma$. This would not be a $7.5\sigma$ detection of the DSNB, but a detection of neutrinos due to both the DSNB and DM. Similar results hold for the mass function of ref.~\cite{Watson:2013}, for a minimum halo mass of $\Mmin=10^{-9}\Msun$, and for a range of DM masses around 20~MeV. 

Interestingly, we find that for $\mdm=30$~MeV the DSNB detection significance actually slightly decreases relative to the no-DM situation. The combined flux of the DSNB and dark matter in this cases becomes more similar to the shape of the background flux within the analysis window, and the likelihood is maximised by assigning some of the signal flux as background. While this effect is quite small and requires the DM mass to be just right, it further demonstrates the importance of quantifying the possible impact of DM on DSNB analyses.

We can also address questions regarding model discrimination. We consider different models $S_i$ of the neutrino signal at HyperK, and calculate a test statistic from the profile log-likelihood ratio 
\begin{equation}
TS= -2 \ln \frac{\mathcal{L}_p \left(\mathcal{D}_A (S_2)|S_1 \right)}{\mathcal{L}_p \left(\mathcal{D}_A (S_1)|S_2 \right)},
\end{equation}
where $\mathcal{D}_A$ is the Asimov dataset~\cite{Cowan:2010js}. The mapping between the test statistic and statistical confidence intervals is not model independent. We assume that the approximate relation $TS\simeq \chi^2$ holds, to enable us to make approximate statements about confidence intervals.

Firstly, we compare between different DSNB models in the absence of dark matter, and without neutron tagging. We consider the low, fiducial and high cosmic star formation histories from Section~\ref{sec:dm}, with $T_{\nu}=4$, 6 or 8~MeV. Taking the first signal $S_1$ to be the fiducial model with $T_{\nu}=6$~MeV, and $S_2$ to be any of the other DSNB flux combinations possible, we find that the maximum TS comparing across all models is 1.7. Thus, although a detection would be made at the $3-4\sigma$ level (see above)  DSNB model-discrimination without neutron tagging would be highly challenging. A more complete recent study on the ability next-generation neutrino experiments to ascertain some of the DSNB parameters can be found in ref.~\cite{Moller:2018kpn}.

Would the presence of dark matter change these conclusions, if there is no neutron tagging? 
Let us now assume that the observed signal $S_1$ is given by the fiducial $T_{\nu}=6$~MeV DSNB model, plus neutrinos resulting from the annihilation of dark matter, $\chi\chi \to \nu\nu$, with $\mdm=20$~MeV. We form the log-likelihood ratio assuming that $S_2$ is one of the DSNB models, in the absence of a neutrino population resulting from DM. In this case we find that the best-fit model is the upper DSNB model with $T_{\nu}=6$~MeV, although this is not statistically significant. However, the lower and fiducial models with $T_{\nu}=4~$MeV would both be incorrectly ruled out at 90\% C.L. by this measurement.

\begin{figure}[t]
    \centering
\includegraphics[width=\textwidth]{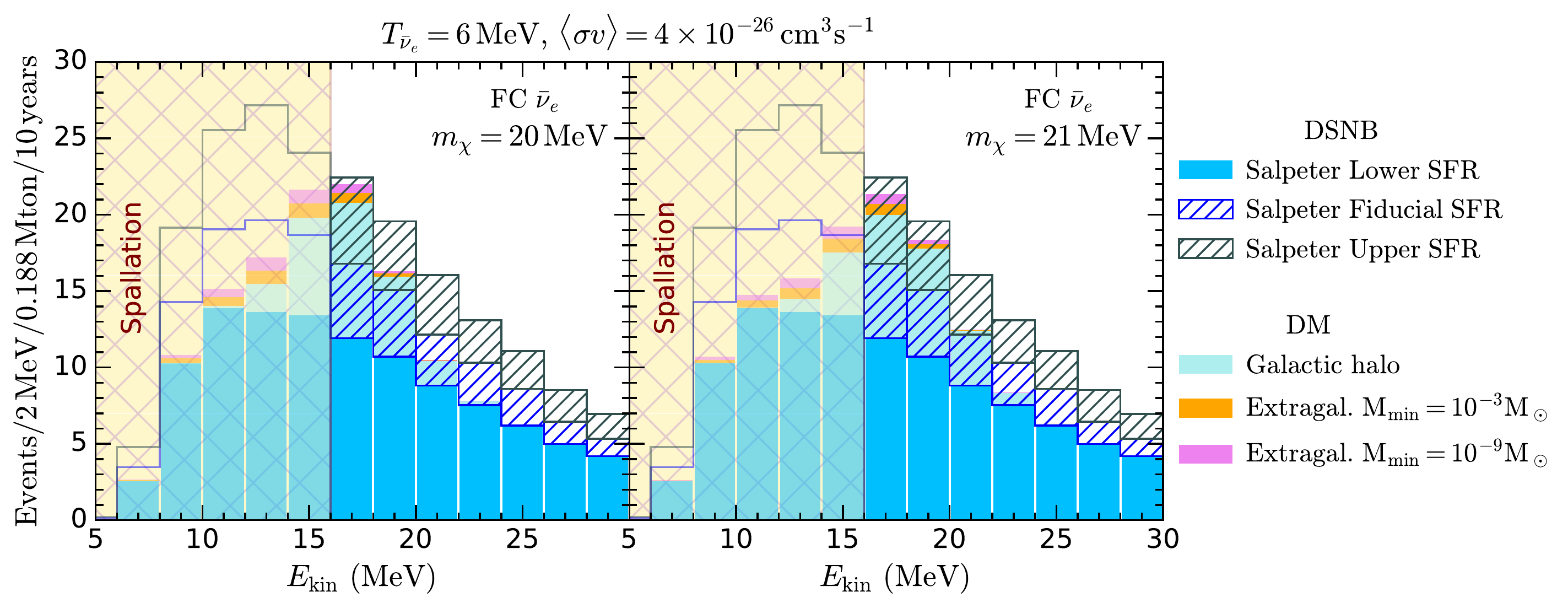}
    \caption{This figure shows the neutrino fluxes resulting from DM annihilating into neutrinos with a thermal cross-section for $\mdm=20$~MeV (left panel) and 21~MeV (right panel), assuming an NFW profile. We divide the DM events to show the galactic and extragalactic components. The latter was calculated using the halo mass function of ref.~\cite{Watson:2013}. We also show the DSNB fluxes for $T_{\nu}=6$~MeV, for the Salpeter upper, fiducial and lower star formation histories. The yellow hatched region below 16 MeV is where spallation backgrounds dominate.}
    \label{fig:dsnb_vs_dm}
\end{figure}

Neutron tagging enables model selection with far greater statistical significance.  For the rest of this section, we assume that the backgrounds are reduced by neutron tagging on Gd as in the HyperK Design Report. As an example, suppose $S_1$ (corresponding to the presumed actual experimental data) was given by the low Salpeter star formation history with $T_{\nu}=4$~MeV, plus neutrinos from 21~MeV dark matter annihilating with a thermal relic cross-seciton. The profile likelihood would lead us to conclude that the low SFR model with $T_{\nu}=6$~MeV  and the high SFR model with $T_{\nu}=4$~MeV give equivalently good best-fits, and the correct DSNB model would be ruled out at over 95\% CL. A neutrino flux originating from dark matter annihilation can thus lead to dramatically incorrect results in DSNB model selection.

We show  some of the relevant fluxes in Fig.~\ref{fig:dsnb_vs_dm} for $\mdm=20$~MeV (left panel) and 21~MeV (right panel), calculated assuming an NFW profile and the HMF of ref.~\cite{Watson:2013}. We show the separate DM components, and the fluxes for the three separate star formation histories. The annihilation cross-section is taken to be $\langle \sigma v\rangle =4\times 10^{-26}\cm^3\s^{-1}$, consistent with the observed relic density for this DM mass. The yellow hatched region shows where the spallation backgrounds dominate below 16~MeV. We find that the effects of DM are maximal for DM masses between 20 and 25~MeV. Below 20~MeV, most of the DM flux is below the analysis threshold and does not affect the DSNB measurement and model inference. For higher DM masses, the difference in shape between the DM and DSNB spectra becomes clearer. For the right mass, the DM flux can add to the neutrino flux in the first few bins above the analysis threshold. This can clearly be seen in Fig.~\ref{fig:dsnb_vs_dm}. The presence of DM with mass 20-25~MeV does not have a large impact on the bins at higher energies. 

\begin{figure}[t]
    \centering
\includegraphics[width=0.7\textwidth]{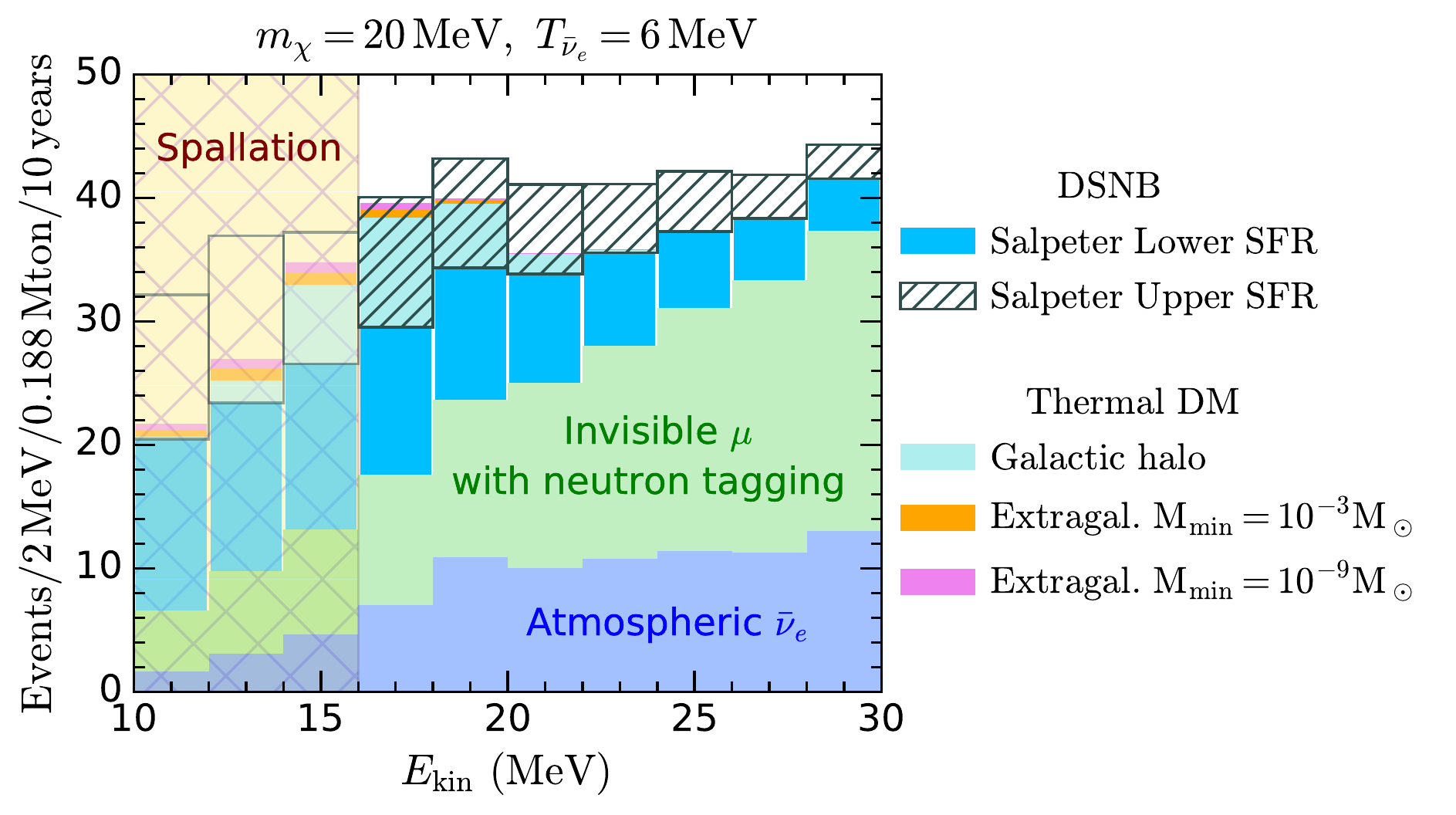}
\includegraphics[width=0.7\textwidth]{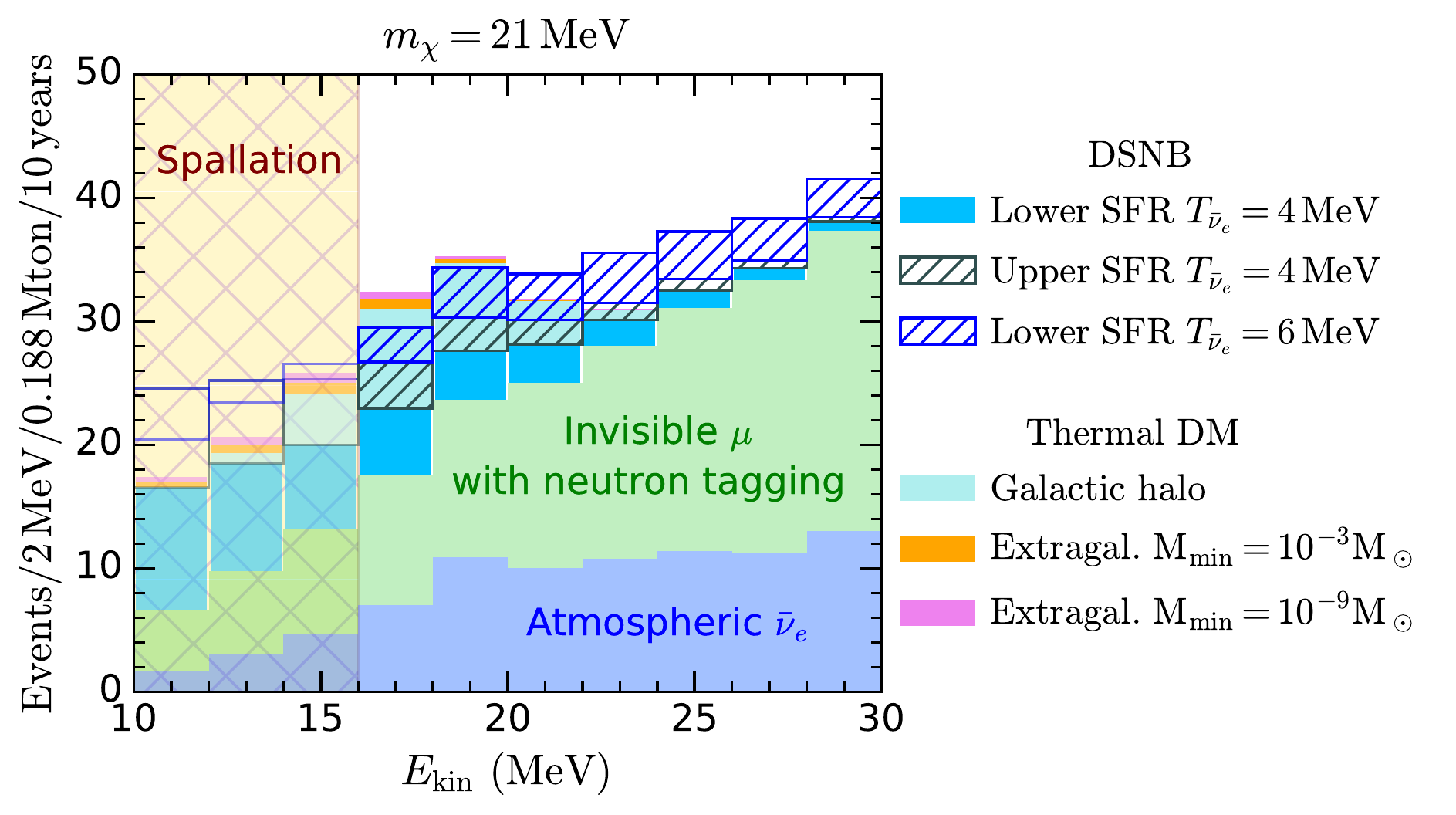}
    \caption{The DM and DSNB fluxes, as well as the backgrounds from atmospheric neutrinos (purple shaded) and invisible muons (green shaded), assuming neutron tagging. (Upper panel): The DSNB data is for $T_{\nu}=6$~MeV and the lower (blue) and upper (green hatched) models. The DM data is for $\mdm=20$~MeV. The presence of DM makes the lower DSNB model resemble the upper DSNB model in the low energy bins where the backgrounds are smallest. (Lower panel):  The DM data is for $\mdm=21$~MeV. We also show the lower 4~MeV (light blue), upper 4~MeV (hatched green) and lower 6~MeV (hatched blue) DSNB models.}
    \label{fig:dsnb_vs_dm_bg}
\end{figure}

However, the backgrounds also increase as the energy increases so that these bins have a lower statistical weight than those at lower energies. We demonstrate this in Fig.~\ref{fig:dsnb_vs_dm_bg}. The upper panel of this figure shows the DM flux for 20~MeV dark matter (light green), and the lower (solid blue) and upper (green hatched) DSNB fluxes with $T_{\nu}=6$~MeV, as described in the paragraph above, but also includes the atmospheric neutrino (purple shaded) and invisible muon (green shaded) backgrounds, including neutron tagging. The lower panel assumes $\mdm=21$~MeV, and shows fluxes for the lower 4~MeV (light blue), upper 4~MeV (hatched green) and lower 6~MeV (hatched blue) DSNB models. 

Given the difference in shape between the DM and DSNB spectra below the 16~MeV spallation threshold, improvements in understanding the spallation backgrounds, which would enable  lowering this threshold, will also be useful in mitigating against DM pollution of the DSNB signal. On the other hand, solar \textit{hep} neutrinos start to become dominant below 18~MeV. The recent JUNO projection study~\cite{JUNO:2022lpc} uses 12~MeV as a lower threshold, and the DUNE technical design report suggests a number around 18~MeV is likely~\cite{DUNE:2020ypp}.

We now discuss the dependence of our results on the dark matter mass and annihilation cross-section. Throughout this section we assume that the dark matter annihilation cross-section is $\sigmav=4\times10^{-26}\rm{cm}^{3}\rm{s}^{-1}$, and standard cosmological history of the Universe. This value yields the correct observed relic density. If the annihilation rate into neutrinos were less than this, dark matter would be over-produced in the early Universe. In principle this could be offset by annihilation into other available channels, such as electrons. However, light dark  matter annihilating into electrons is tightly constrained by CMB measurements~\cite{Slatyer:2015jla}. On the other hand, annihilation cross-sections  larger than $\sigmav=4\times10^{-26}\rm{cm}^{3}\rm{s}^{-1}$ lead to underproduction of dark matter in the early Universe, which would thus constitute a sub-component of the total relic density. The increase in $\sigmav$ in Eq.~\ref{eq:dmflux} is offset by the decrease in the J-factor, which is proportional to $1/\sigmav^2$, and thus the net result is a smaller  annihilation flux. A larger annihilation cross-section thus kills off the signal, assuming a standard cosmological history.

Dark matter which can impact DSNB analyses must lie in a specific mass range, between 17 and 30~MeV. This is determined by the size of the DSNB analysis window. However, thermal dark matter at the upper end of this range would only have a small impact on DSNB analyses. This is because the backgrounds are larger at higher neutrino energies, and because the neutrino fluxes from DM annihilation, given in Eq.~\ref{eq:dmflux}, are proportional to $1/\mdm^2$. Accordingly, we approximately delineate the DM mass-range relevant for this effect as 17 -- 25 MeV.

Can we disentangle such a dark matter contribution from the DSNB signal? Angular information would be highly useful, because the annihilation signal is dominated by the Galactic centre contribution, while the DSNB is isotropic. Unfortunately, the dominant detection channel in the relevant low energy range, namely inverse beta-decay (IBD), has very weak directionality.    
Around the DSNB energy window, there is a cancellation in the angular differential IBD cross-section which renders it very nearly isotropic in the angle between the incident anti-neutrino and outgoing positron energy~\cite{Vogel:1999zy}. Consequently, it is not possible to reconstruct the direction of the incident anti-neutrino. Thus, although the majority of the dark matter annihilation flux originates from the direction of the Galactic centre, this information is not accessible at HyperK. While better directional reconstruction is possible with electron-neutrino scattering, the cross-section is suppressed relative to that for IBD by a factor of around 50 at the energies we consider~\cite{Tomas:2003xn}. Given that we are talking about $\mathcal{O}(10^2)$ events, the electron-neutrino scattering channel will not be of sufficiently high statistics to be useful.

%%%%%%%%%%%%%%%%%%%%%%%%%%%%%%%%%%%%%%%%%%
\section{Summary}
\label{sec:summary}
%%%%%%%%%%%%%%%%%%%%%%%%%%%%%%%%%%%%%%%%%%5

The discovery of the diffuse supernova neutrino background (DSNB) and dark matter (DM) are eagerly anticipated. Upcoming neutrino experiments may have sensitivity to both. We have shown that light dark matter which annihilates into neutrinos can lead to a flux of neutrinos in the energy range where the DSNB is present. This ``pollution" from neutrinos with a DM origin could confound precise measurements of the DSNB, leading to incorrect inferences about the astrophysics behind the DSNB, and potentially missing a signal due to DM. We have demonstrated this via the use of simulations of the Hyper-Kamiokande (HyperK) detector, building on our previous work on DM searches in the Galactic centre and the Sun.

However, we have argued that it will be difficult to discriminate between dark matter and the DSNB at HyperK due to the  isotropic nature of the predominant inverse beta-decay events and small overall event numbers. While we have focussed on HyperK in this paper, our conclusions should hold for any experiment which has sensitivity to the DSNB, including DUNE and JUNO. We therefore advocate that future DSNB analyses consider dark matter annihilation to neutrinos as a possible background in the DSNB energy window.

\section*{Acknowledgements}
NFB and MJD were supported by the Australian Research Council through Discovery Project DP220101727 and through the ARC Centre of Excellence for Dark Matter Particle Physics, CE200100008. SR acknowledges support from the UK STFC grant ST/T000759/1. 
We thank Teppei Katori and the anonymous referee for helpful comments.

%%%%%%%%%%%%%%%%%%%%%%%%%%%%%%%%%%%%%%%%%%%%%%%%%%%%%%%%%%%%%%%%5
\bibliographystyle{JHEP} 
\bibliography{references}

\end{document}